# A review of assessment methods for the urban environment and its energy sustainability to guarantee climate adaptation of future cities

Dasaraden Mauree[a], Emanuele Naboni[b], Silvia Coccolo[a], A.T.D. Perera[a], Vahid Nik[c,d,e], Jean-Louis Scartezzini[a]

[a] Solar Energy and Building Physics Laboratory, Ecole Polytechnique Fédérale de Lausanne, CH-1015, Lausanne, Switzerland
[b] Institute of Architecture and Technology, The Royal Danish Academy of Fine Arts, Schools of Architecture, Design and Conservation, 1425 Copenhagen, Denmark
[c] Division of Building Physics, Department of Building and Environmental Technology, Lund University, 223 63, Lund, Sweden (nik.vahid.m@gmail.com, vahid.nik@byggtek.lth.se)
[d] Division of Building Technology, Department of Civil and Environmental Engineering, Chalmers University of Technology, 412 96, Gothenburg, Sweden
[e] Institute for Future Environments, Queensland University of Technology, Garden Point Campus, 2 George Street, Brisbane, QLD, 4000, Australia

## Contribution Highlights

- Reviewed assessment methods for the urban environment
- Critically analysed papers working on urban climate and energy demand, outdoor thermal comfort and the urban energy systems.
- Demonstrated the links between the processes
- An integrated workflow is proposed for assessment of the urban environment.

## Abstract

The current climate change is calling for a drastic reduction of energy demand as well as of greenhouse gases. Besides this, cities also need to adapt to face the challenges related to climate change. Cities, with their complex urban texture and fabric, can be represented as a diverse ecosystem that does not have a clear and defined boundary. Multiple software tools that have been developed, in recent years, for assessment of urban climate, building energy demand, the outdoor thermal comfort and the energy systems. In this review, we, however, noted that these tools often address only one or two of these urban planning aspects. There is nonetheless an intricate link between them. For instance, the outdoor comfort assessment has shown that there is a strong link between biometeorology and architecture and urban climate. Additionally, to address the challenges of the energy transition, there will be a convergence of the energy needs in the future with an energy nexus regrouping the energy demand of urban areas. It is also highlighted that the uncertainty related to future climatic data makes urban adaptation and mitigation strategies complex to implement and to design given the lack of a comprehensive framework. We thus conclude by suggesting the need for a holistic interface to take into account this multi-dimensional problem. With the help of such a platform, a positive loop in urban design can be initiated leading to the development of low carbon cities and/or with the use of blue and green infrastructure to have a positive impact on the mitigation and adaptation strategies.

## Section 1: Introduction

According to the 5[th] assessment report on climate change from the Intergovernmental Panel on Climate Change (IPCC) [1], there is no doubt that anthropogenic greenhouse gas emissions (GHG) are responsible for the current climate change. The recent special report on the impacts of global warming of 1.5°C [2] was yet another call to implement measures to mitigate GHG emissions and also to devise new adaptation scenarios. Thus, if the Paris agreement is the objective, there is an urgent need to reduce our energy demand and decrease GHG emissions. Additionally, with an inevitable 1.5°C increase in global temperatures, adaptation strategies to improve the design of urban areas (more liveable spaces) and energy systems are required.

Around 3.5 billion people live in urban areas around the world and by 2050 more than two-thirds of the urban population will live in cities [3]. Around two-thirds of global primary energy demand is attributed to urban areas, inducing 71% of global direct energy-related GHG emissions [4]. The combination of the projected population and economic growth together with climate change results in placing greater stress on vital resources in the future if there is a continuation of the business as usual scenario [5]. The energy sector in urban areas could thus play an important role to tackle climate change and to decrease the carbon/energy footprint of urban areas.

Besides this, urban development has also lead to the Urban Heat Island phenomena (UHI) [6] which causes a significant increase in air temperatures in urban areas and are hence exacerbating the effects of climate change with the increase in heat waves in the future [1], [2]. For example, it is predicted that in the RCP 8.5 scenario, there will be up to 17 (30) more days of tropical nights by 2060 (2100) in Switzerland [7]. It is thus evident that there is a need to increase the comfort and the design of buildings to adapt in the most comprehensive way to the negative impacts of climate change [8]. A few cities are already taking adaptive actions and cities are seen as the 'first responders' to climate change. The measures taken presently are mostly related to flooding or storm surges and the liveability and future energy demand of the urban space have not been particularly dealt with by the urban planners [9]. It is thus clear that there is a willingness to "control" the effects of climate change either from the emissions perspective or from the adaptation one, in particular when looking at urban areas. The following actions are often cited as necessary measures in the transition in particular in mid-latitudes countries but are often contradictory among themselves:

1. Reduction of the energy demand for the operation of the buildings without consideration for the urban texture and fabric.
2. Promotion of outdoor comfort strategies to mitigate the significant overheating during summertime without consideration for urban planning strategies
3. Develop low carbon energy systems without or with little consideration for the energy efficiency of the urban system (which could lead to an increase in energy demand for example with the rebound effect).
4. Integrate renewable energy without or with little consideration for the urban texture or fabric.

However, most of the time, the climatic challenges (e.g. comfort, energy demand, energy systems…) are assessed individually while they are likely to be interrelated and require a holistic understanding of the ecosystem and human activities and the built environment (its form and fabric) [10] that could lead to a regeneration of the urban space. Designing a single, often free-standing, low carbon building is different from planning an urban area. Several studies combine urban fabric ancient knowledge with fast computing techniques, virtually showing that low carbon cities are a realistic option [11]–[13]. However, the current practice of building design was, for years, shaped by building codes with little use of such knowledge and modelling capabilities that have been marginalised from the design. A network approach, based on the modelling of the climate, city, buildings, outdoor spaces and human variables, is necessary to understand their interdependencies [14] and make decisions that impact each of these components positively. Embracing the complexity of such networks by modelling it, under both the typical conditions and the extreme climatic events of thermal peaks, is today key [15]. This adds complexity to the design process, and it is clear that specific design methods and tools, able to model, simulate and assess, need to be adopted by designers and urban planning specialists. The use of specific modelling tools will allow for an adaptive design truly resilient to changes and decrease uncertainty in a way that would make buildings better [16].

Multiple studies have recently used Integrated Assessment Models (IAMs) at the global scale to analyse and forecast the implications of climate change on socio-economic variables [17], [18]. However, they also noted that these tools often lack the precision to give realistic indications at the urban scale. Previous reviews have also been conducted but remained rather focused on one single aspect. For instance, Nault et al. [19], looked at the evaluation metrics to assess solar potential in an early design phase. Haapio et al. [20], emphasised on the building environmental assessment tools while others have focused on the impacts of urban energy systems[21]. Some recent studies have also given some insights into the type of infrastructure that need to be developed in order to provide useful information to urban planners [22]. Keirstead et al. [21] already demonstrated that there was a need to move beyond "single disciplinary approaches towards a sophisticated integrated perspective that more fully captures the theoretical intricacy of urban energy systems".

There is thus a lack of a comprehensive review that focuses on the interrelation between the energy demand, urban microclimatic conditions, the energy systems optimisation, the outdoor thermal comfort, as well as the future climatic conditions . We have hence analysed the existing cohort of studies on assessment tools used to evaluate these aspects focusing on the urban environment and energy sustainability. In this review paper, it is proposed to evaluate the types of interdependencies that exist between the four targets, to determine which assessment tools are available to address them and further evaluate future climate adaptation targets for cities and to define how they account for the conflicting aims. The objective is therefore to highlight the most common criterions (or key performance indicators) used in these assessment tools and their interdependencies, to provide an overview of assessment tools and to propose, finally, a holistic approach.

The paper is structured in the following way: first, we explain the methodology used to conduct this review. Then we describe the processes taking place in the urban space and that will be relevant to the urban climate, the outdoor thermal comfort, the energy demand and the energy systems. Third, we analyse the tools used in their assessment. The performance of the different tools in particular in the context of climate change is also considered. Some of the tools that have been brought together in one framework are also discussed. Finally, we give some perspectives and how the limitations of the current tools could be addressed in the future.

## Section 2: Review methodology

The cohort of papers was selected using the following keywords: "urban climate", "urban heat island", "urban energy demand", "urban energy systems", "outdoor thermal comfort" and "climate change". Among all the papers that were obtained, 187 papers were chosen based on their relation to the study. **Figure 1** shows the significant increase in the number of articles published since the 1960s as a reference by Scopus along with the evolution of the selected papers. **Figure 2** gives a categorical breakdown of the papers obtained from the Web of Science.

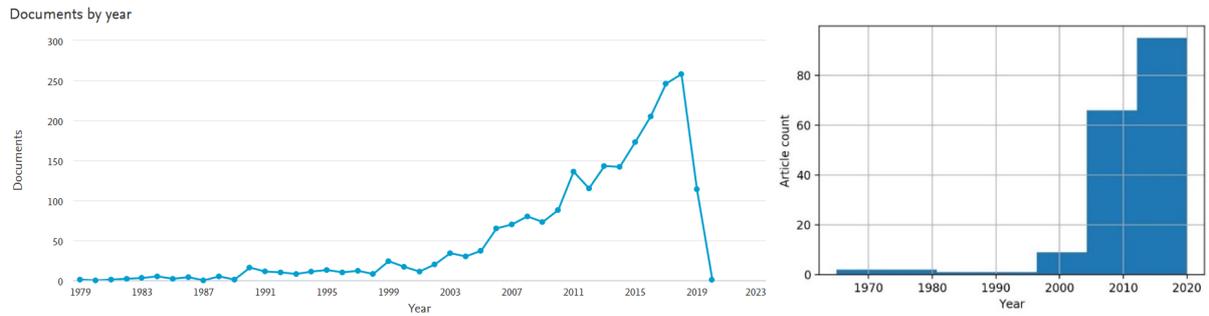

*Figure 1: Number of papers published using all keywords from Scopus (left) and the one selected for the review (right).*

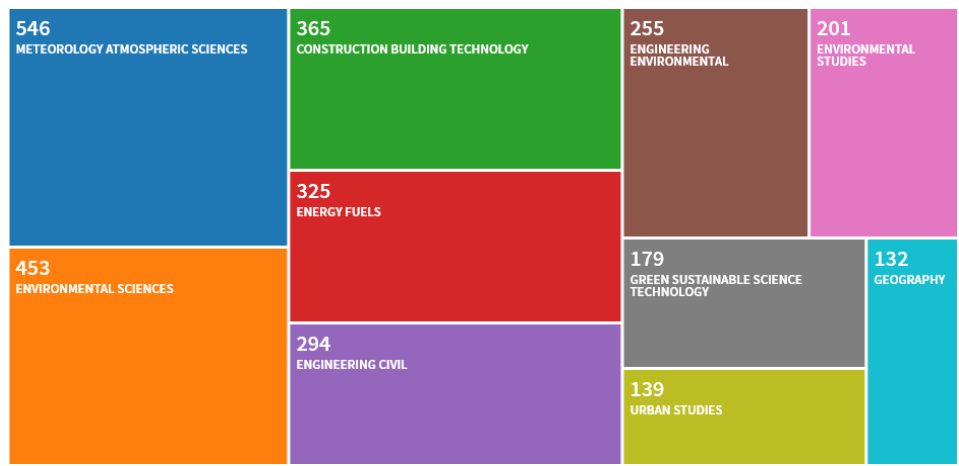

*Figure 2: Categorisation of all papers, using the keywords, obtained using Web of Science.*

An analysis of the existing associations between the keywords from these papers was conducted with the Voyant Tools [23] (see **Figure 3**). It can be noted from this analysis that although there is an obvious predominance of the links between the words "urban", "energy" and "climate", connections exist between multiple other keywords and which to the authors knowledge have not yet in addressed in a review.

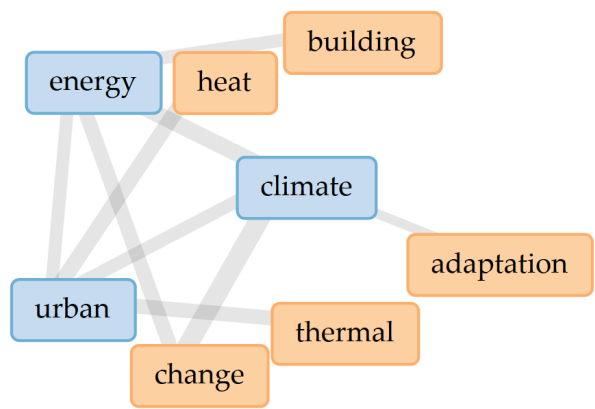

*Figure 3: Analysis of the keywords from the selected papers, from Voyant Tools [23]. Boxes in blue represent the dominant terms while boxes in orange represent the connected terms*

It can be highlighted that when using all the keywords mentioned before, no paper was referenced in either Scopus or Web of Science and only one review paper from Keirstead et al., [21] was obtained in Google Scholar. This points to the need for more research in this area to understand the implications of future climate change on the urban design not only from a building perspective but also the impact on the energy systems as well as on the thermal comfort. In this paper, we will give thus look in detail at the three primary keywords (urban, energy and climate) to understand these links and in order to build a framework that could address them as well as the connecting keywords with the objective of designing more sustainable urban areas.

# Section 3: The urban space

## Section 3.1 Urban Climate

Given that buildings are responsible for 40% or more of most countries' GHG emissions that contribute to climate change [24], strategies on how to reduce their energy demand are an integral component of the urban design. However, cities need to offer their residents healthy and attractive indoor and outdoor spaces. An enjoyable microclimate is critical but is often undermined by the current urban planning which is seldom attempting to tackle climate change. Furthermore, buildings operations and outdoor environment are thermodynamically looped.

Urban microclimates are both complex and dynamic, and they hold many profound implications for successful urban planning and building design. Early research generated insightful discoveries which provided evidence on how building and open spaces form affect the urban climate [25]–[27]. It is thus clear that buildings and spaces together lead to a specific microclimate, that is different from the rural sides [28]. Compared to open country, built urban sites have more areas of exposed surfaces per unit area of ground cover[29]. Because of these large areas, more solar radiation can be collected on a built urban site than on flat, open terrain, with implications on the microclimate. In the city, a surface's exposure to the sun and wind at any given time is mostly determined by the built form, as well as the street widths and orientation.

According to Oke [30], the UHI results from the combination of the phenomena mentioned above, which generally increase urban surfaces temperatures, in addition to the high thermal absorbance of urban materials, the lack of vegetation (evaporative cooling), and the anthropogenic heat sources. Buildings operations lead to surface temperatures that are not only given by sun radiation but also due to internal activities and mechanical system operations. Heating, ventilation and air conditioning (HVAC) systems, finally, often release exhaust heat in streets contributing to modify the local microclimate and potentially leading to a loop where they have to respond to their effects on the local microclimate. Taking as reference countries with hot-arid and tropical climate a vast amount of energy, particularly electricity, is consumed for cooling buildings [31]–[33] and they exhaust significant heat in the urban fabric. This negative loop is amplified as conditions worsen with climate change and induce warmer weather on average as well as stronger and more frequent extreme conditions [34], [35] increasing the average and peak cooling load as well as thermal discomfort in urban areas with hot summers [15], [36].

## Section 3.2 Urban Fabric

In the past, common sense and tacit knowledge have led to a settlement where building operations and outdoor spaces were very finely integrated contributing to spaces within the building walls and on the outdoor. An example of the above is the city of Shibam in Yemen and the city of Fez in Morocco [37], where the urban design itself autoregulates the outdoor environmental conditions. This and other examples show that the urban outdoor space could be designed to create a microclimate that supports single buildings indoor comfort and thus energy savings. Vernacular examples show that buildings and urban microclimates are interwoven: urban microclimates affect a building's energy demand (and indoor environment), while buildings affect the urban microclimate [38]–[41]. As noted by Givoni [42]: "The outdoor temperature, wind speed and solar radiation to which an individual building is exposed is not the regional 'synoptic' climate, but the local microclimate as modified by the 'structure' of the city, mainly of the neighbourhood where the building is located". These modifications lead to significant modification of the building energy demand[11], [43], [44].

However, this knowledge was lost in recent urban design: as cities development lead to a substantial increase in the gross floor area (GFA) (and will continue to do so [45]), modified aspect ratio (AR) and the hardscape replaced the softscape further amplifying this loop inducing warmer temperatures in cities than in the surrounding areas, leading to the urban heat island effect [46]–[48]. The impact of the air temperature increase in urban areas can lead to higher demands for air-conditioning and cooling in more extended periods, compared to rural areas [49], [50]. Moreover, the UHI is logarithmically proportional to the population [51]. Within the context of cities growing in population, this effect is thus increasing. This behaviour is observed in numerous field studies around the world for a variety of climatic regions [52]. For example, Kalnay and Cai [53] have shown that rapid urbanisation and high urban density have caused 0.27°C mean air temperature increase during the last decade.

## Section 3.3: Outdoor thermal comfort

As discussed above, it was observed for a long time that cities alter the local climate. Indeed, air temperature and wind speed patterns and species concentrations (such as humidity or other air pollutants) are modified compared to rural areas. This in return impacts the building energy needs and the liveability of the urban space.

Outdoor thermal comfort plays a significant role in urban sustainability, directly affecting people's health and wellbeing. Due to rapid and intensified urbanisation trends, new attention to the conditions of comfort and liveability of our cities is given. It is demonstrated that people's thermal comfort is one of the factors that affect the fruition of urban spaces like streets, plazas and parks [54]–[56]. The quality and successful usage of these spaces can have further implications for the development of our cities. Hence, understanding and evaluating thermal comfort conditions in urban spaces is necessary. It is today widely investigated how the built environment can alter local microclimates by influencing a series of thermodynamic phenomena, which affect substantially human thermal comfort conditions [57]. Addressing outdoor comfort conditions involves issues not encountered in indoor comfort studies [58]. Pedestrians may be exposed to different solar radiations and wind speeds, which may greatly vary in time and space, thus affecting their perception of the surrounding environment [56].

When working on the outdoor thermal comfort, it is essential to understand how the pedestrian lives the space, what is their physical, psychological and physiological adaptation to the environment [59], [60]. Indeed, as a function of their motivation of being in the space and the visual clues, the thermal response and reaction to the environmental factors vary importantly [61]–[63]. In this context, urban surfaces play a crucial role in mitigating urban environmental conditions, consequently improving the pedestrian's health and well-being [64], [65]. As an example, the use of reflective surfaces [66], cooling materials [67] and greening [68], [69] can significantly improve the environmental urban conditions, reducing the radiant temperature, improving the natural ventilation and mitigating the urban heat island effect. It is now common knowledge that the ground thermal properties impact the radiation (longwave and shortwave) absorbed by the pedestrians, mostly due to its colour and thermal mass [70], [71]. Also, the thermophysical properties of the facades of the buildings play a significant role in the urban comfort, since they can reflect, or absorb, the solar radiation, consequently directly impacting the city liveability. Extending the scopes of the debate, the city form itself impacts not only buildings energy demand and thermal comfort; they affect our perception of the space and human cognition [72], [73].

When focusing on the population's health, it is evident that a "nice" urban design can positively affect the citizen's health and well-being. Indeed, during the more frequent summer heat waves, the vulnerable population, such as older people and children, is the one suffering the most [74], [75]. Consequently, it is essential to develop models able to understand and predict their thermal sensation, in order to prevent the consequences related to heat stress. As an example, it is currently difficult to quantify the children's thermal sensation, due to their physical and physiological reactions, which are different than in adults. Some recent works have started to address this question to understand and quantify their thermal behaviour [76], [77]. Within these problems, the urban planning design plays a significant role and should consider the impact of the city design on the thermal comfort and health of the population, providing concrete solutions for sustainable and comfortable design, able to face the climate change [78].

## Section 3.4: Urban energy demand and systems

The general UHI effect, which means that air temperature within a city is often higher than in rural areas, decrease the heating needs but increase the cooling one. Several studies additionally showed that the UHI leads to a reduction of around 10% in buildings energy demand in cold climates and to an increase of around 20% of cooling needs in tropical climates [43], [52], [79]. At mid-latitudes as a result of the UHI, the demand is reduced by 25% in heating and increased by 15% in cooling [80]–[82]. A recent work has assessed the impacts of urban morphology on reducing cooling demand and increasing ventilation potential in hot-arid climates, investigating the effects of urban density, urban building form and urban pattern [33].

Major trends proposed to reduce the energy demand of the existing building stock is the improvement of the thermal performance of single buildings. A series of norms and standards have pushed toward new construction standards to drastically minimise the energy demand of new and retrofitted buildings and minimise the associated greenhouse gas emissions [83]. In the literature, the Zero Energy objective is mostly considered at the building scale [84]. Several papers have proposed definitions of Zero Energy Buildings, calculation methodologies or support tool for early stages of design considering the individual building as an autonomous entity [85]–[87] and neglect the importance of the thermodynamic link the building with the outdoor microclimates, thus neglecting to model heat and mass flow in and around buildings. The neighbourhood scale is relevant from an operational point of view and allows to take into account thermodynamic interactions that occur at an urban scale. Furthermore, the impacts of parameters linked to the urban form on the energy demand of single buildings and the efficiency of renewable energy sources are considered key [88]–[90].

The energy demand represents the energy used by energy systems, considering their efficiency and their behaviour, to provide energy services. The energy demand refers to the assessment (the sum) of the energy demand over a period. Hourly energy demand is commonly used in building energy simulations as the minimum temporal resolution required to estimate the power demand [91]. Simulating urban building energy demand is more complicated at the city scale than at building scale, due to the significant amount of data related to the built environment and user behaviour [21], [92]. Effects of the urban microclimate and the surrounding spaces and buildings need to be accounted for, while external loads, such as meteorological loads, cannot be estimated so generically as they are typically simulated in single building energy simulation.

More precisely regarding this last point, meteorological loads of urban buildings and subsequently their energy behavior depend on obstructions caused by surrounding constructions, which decrease the sky view factor, and consequently reduce solar gains (increase of the heating needs in winter and decrease of the cooling needs in summer) and the radiative cooling to the sky (reverse effect on the space conditioning needs). Surrounding surfaces, which reflect solar radiations and emit and reflect longwave radiations, impact on the surface energy balance of urban buildings (e.g. a north-oriented surface may receive solar radiations from a south-facing opposite surface [93]. Therefore its thermal losses may be reduced. Furthermore, the urban morphology modifies airflows around buildings, and, consequently, impacts convective heat exchanges [94], [95] and the potential of natural ventilation of urban buildings, including infiltration [96]. In recent years, significant progress has been made towards the development of simulation workflows to estimate overall operational building energy use but have often been limited to a few neighbourhoods [92].

Quantifying global solar irradiation hitting building envelopes and assessing the potential for photovoltaic electricity production and solar thermal for space/water heating (active systems) and solar heating (passive systems) have also received much attention in the past decade. While active solar systems use mechanical and electrical devices to convert solar radiation to heat and electric power, passive solar design uses building design to collect and harvest the sun's heat and to reduce the energy required for space heating. It is thus clear that renewable energy integration in urban areas, need to account for the external environmental conditions.

Effects of urban climate and climate change are not limited to the demand side and are extendable to energy systems and infrastructure, as it has been investigated for some cases in the USA [97], Greece [98], Norway [99] and Australia [100]. Climate change (and its uncertainties) can affect renewable energy generation (and its estimation), especially wind [101], hydropower [102] and solar energy [103]. Decentralised on-site energy production and use in urban areas are expected to minimise the loss or transformation energy transmission. The integration of renewable energy in urban areas will provide new opportunities for urban energy system [104]. In particular, the resource that has grown the most in the last decade is solar energy [105], and it is very likely that it will provide the largest share of the electricity mix by 2050 [105]. Based on the current growth rate (with continued policies and technological progress), by 2050, solar photovoltaics and solar thermal energy could contribute to 27% of the global electricity production. The local production of energy from renewable resources is expected to contribute to a significant reduction in greenhouse gas emissions from the production using fossil fuels.

Moreover, the impact of energy systems will depend on multiple spatiotemporal factors. Although the impact of climate change will be minimal for some resources such as wind [99], this will neither be necessarily the case for the operation of energy systems or power plants nor on the demand side. Shen and Lior [106] also assessed the sensitivity of climate change on the performance indicators of a net-zero energy building (considering both demand and generation) obtained using a deterministic model. Dowling [107] evaluated the impact of climate change on the energy system at a regional scale considering entire Europe and based on several scenarios (although design optimisation is not considered in this work). Finally, Mavromatidis et-al [108] demonstrated that climate change along with the uncertainty due to occupancy and cost of energy technology will have an effect on the optimum energy system design.

# Section 4: Assessment of the urban built environment, urban thermal comfort and energy systems

Climate change is referred as changes in statistical distribution patterns of climate variables which leads to more extreme and more frequent weather events such as heat waves that have a notable impact on the energy infrastructure [1]. It is a difficult task to quantifying the risk introduced by weather due to its high stochasticity and multi-dimensional impact [109]. Climate change will affect both the energy demand and the supply, for example increasing the cooling demand and decreasing the heating demand [110] and intensifying extreme events [15], threatening the security of generation, transmission, and distribution infrastructure [111]. A number of studies have focused on the impact of climate change at both national [112], [113] and continental scale [114], [115].

Energy transition in the urban sector should address objectives such as improving sustainability, reduction of noxious gases, such as $SO_x$, $NO_x$ and particulate matter due to conventional power generation, improve the efficiency of energy conversion methods, reduce the cost related to power generation, improving the reliability and security of power supply [116]. Most of these objectives are directly connected to climate change while some others are indirectly related while covering a broad spectrum of requirements in society. However, linking these objectives with the energy system designing process is a challenging task. As a result, assessment of energy infrastructure has been often performed on a sectoral basis without much coordination among each [117].

In the previous section, we have described the processes influencing the urban climate, the outdoor thermal comfort, the urban fabric and the urban energy systems. For a complex space such as urban areas, modelling tools can be used as a means to assess different strategies for the urban space [118]. Criterions can thus be used but are highly subjective to the design requirements which will depend on various factors such as customer requirement, topography, financial constraints etc. Hence, criterions suggested in this study only provide an outline for selection of criterions for assessment and is a non-exhaustive list of criterions that could be used as it is or modified when applying to specific cases. In the next section, it is described how models have been used to provide key performance indicators to assess the urban climate, the energy demand, outdoor comfort and to design the energy systems.

## Section 4.1: Urban climate

As mentioned in Section 2, there is a close and intricate relationship between the urban climate and the energy demand, the outdoor comfort and the energy systems. The modelling of urban microclimates is very complicated because of the cities' geometric complexity and heterogeneity as well as the relation to atmospheric phenomena. The use of the governing equation of fluid dynamics takes into account the strong interactions between buildings and microclimate and require the use of coupled approaches which will be explained later in the paper. Three critical meteorological variables are often computed to evaluate the urban climate: the air temperature, the wind speed and the humidity (see **Table 1**). Other variables such as the sensible heat flux, the latent heat flux, the surface temperature or the air pollutants concentrations are also regularly used.

Locally measured data can also be specified as inputs for the urban building energy model. However, collecting suited measurements necessitate expensive and extensive experimental field campaigns, which are necessarily limited, and which can only be set to the current location. To overcome this limitation, it is possible to extend measured weather data from one place to another place thanks to extrapolation techniques. This is the case, for example, with the urban canopy models or computational fluid dynamics (CFD) models which can provide high-resolution data. Although CFD models are very useful, in understanding local-scale phenomena and typically resolve the flows with a very high resolution, it can be noted, nonetheless that CFD models cannot be used for an area larger than a district and simulations cannot be run with an hourly time step for a full year. This significantly limits their usage in the assessment of urban planning scenarios. A detailed review of CFD models for use in urban area can be found in [119]. Thus tools such as the Canopy Interface Model (CIM) [120] have been developed with the aim of improving the land-surface processes in climatic models. CIM contains a parameterisation similar to the Building Effect Parameterization [121] but improved with a vertical diffusion process. Other microclimatic tools have also been developed in recent years. Two models currently used are the Urban Weather Generator (UWG) [122] and Canyon Air

Temperature (CAT) [123]. Their starting point is a measurement point outside of the city under the same mesoscale climatic conditions, like an operational measure station at an airport or standardised weather [124]. UWG is based on energy conservation principles and is a bottom-up building stock model. The building parameterisation is similar to the Town Energy Balance model [125]. Based on its inputs, it computes a rural profile and then uses an urban boundary layer model to obtain air temperature values for the urban site. It can be extended for simulations at the city-scale. A spatial urban weather generator (SUWG) that calculates a 2D (horizontally) field for the temperature above the urban canopy layer was further developed. The CAT model can be used in order to simulate the air temperature in a specific site of a city. However, due to the lack of advection processes, it cannot be used to simulate local city-scale UHI. Contrary to UWG and CAT, CIM also computes the horizontal wind speed and has recently been extended and integrated into the Weather Research and Forecast model [126] to include advection processes [39]. It was shown here that there was a significant improvement to the computation of high-resolution vertical profiles which can play a significant role in the computation of the building energy demand at the urban scale [127] and the outdoor thermal comfort [128].

*Table 1 : Selection of variables used for the urban climate*

| Main variables | Other variables |
|---|---|
| - Air temperature (°C or K) | - Sensible heat flux (W/m$^2$) |
| - Wind speed (m/s) | - Latent heat flux (W/m$^2$) |
| - Relative humidity (%) | - Surface temperature (°C or K) |

## Section 4.2: Outdoor thermal comfort

Design choices change urban environments by changing the thermodynamic phenomena, which consequently alter human thermal comfort. This makes it imperative to focus on microclimatic design to raise people's health and wellbeing [129]. Because of the dynamic nature of the urban environment, it is still difficult to quantify and manage the physical variables that play a role in urban microclimates [130]. To understand the importance of modifying the outdoor climate in a particular direction by specific design choices, several comfort indexes and physical parameters have been introduced in the evaluation of the comfort conditions of persons staying outdoor. Additionally, when addressing the outdoor thermal comfort, it is always quite difficult to interconnect architectural needs with the rigorous biometeorological protocols.

There are several methods of determining the quality of outdoor microclimates [131], one of them is the use of biometeorological indices (see **Table 2**), allowing the quantification of thermal comfort as well as heat [132]. In the last decade, the scientific community interests in outdoor comfort lead to modelling tools able to predict microclimatic conditions [133]. But, it is currently quite a challenge to model the outdoor comfort quality of design options [134], [135] although, urban designers are aware of the importance of the local microclimate [136].

The potential users are confronted with the dilemma of choosing a suitable outdoor comfort simulation tool. Indeed, architects and urban planners are challenged with the selection of tools that fit in their "digital ecosystems": these being either GIS/BIM-based or Rhino/Grasshopper ecosystem[137]. While the integration of building energy simulation into the design process is mainly achieved [138], this is not the case for the outdoor microclimatic simulation. Additionally, outdoor climatic simulation tools developers rarely state the tool's capabilities and limitations but some of their capabilities have been assessed in previous research [139].

Several tools exist (CitySim [130], RayMan [140], ENVI-met [141], SOlar and Long Wave Environmental Irradiance Geometry (SOLWEIG) [142], Grasshopper plug-ins Honeybee and Ladybug [143]) to understand and to model the outdoor thermal comfort and are well validated and used within the scientific community. CitySim simulates and optimises urban settlements by predicting energy fluxes at various scales, from a small neighbourhood up to a small city. With its microclimatic modelling, it is possible to quantify the Mean Radiant Temperature (MRT) [133], the Index of Thermal Stress (ITS) [78] and the COMFA* budget [144]. Additionally, a first interconnection between the tools CitySim and RayMan was proposed, showing a good agreement between the two software [145]. RayMan is an established tool to compute the outdoor thermal comfort and is widely used all around the world. It computes the radiation fluxes and thermo-physiologically indices as the Predicted Mean Vote (PMV), the Physiologically Equivalent Temperature (PET), the Standard Effective Temperature (SET*) [146], the Universal Thermal Climate Index (UTCI) [147], Perceived Temperature (PT) [148] and the MRT.

SOLWEIG, also calculates the PET, the UTCI and the MRT for complex urban settings [142]. ENVI-met [141] is a three-dimensional microclimate model designed to simulate the surface-plant-air interactions in an urban environment by defining the microclimatic conditions of the selected sites. The plugin BioMet calculates the following indices: PMV, PET, UTCI and MRT. Ladybug and Honeybee [143] are two open source environmental plugins for Grasshopper built on top of several validated simulation engines. They integrate the outdoor thermal comfort into the design flow and can thus be an essential instrument to bring awareness to architects and urban planners. They indicate the outdoor environmental conditions, as well the pedestrians' thermal sensation, through several indices such as the PET and the UTCI. They have been applied to compute climatic conditions that range from urban canyons to city scale [149], [150]

*Table 2 : Selection of indices used for outdoor thermal comfort*

| Temperature based indices | Other indices |
|---|---|
| • Mean Radiant Temperature (MRT) | • Index of Thermal Stress (ITS) |
| • Perceived Temperature (PT) | • Universal Thermal Comfort Index (UTCI) |
| • Physiologically Equivalent Temperature (PET) | • Predicted Mean Vote (PMV) |
| • Standard Effective Temperature (SET*) | • COMFA* |

### Section 4.3: Energy demand and supply

There are complex energy flows at the urban scale that needs to be accounted for in the computation of energy demand in urban design and to support decision-making. Whereas single buildings modelling was widely developed and are now used in practice, the urban energy demand modelling is a relatively new field. As the objective of this review is also to look at the impact of future climatic changes and of planning strategies, we did not look at statistical approaches to compute the energy demand. For these described reasons, only deterministic bottom-up tools that reconstitute the behaviour of a city from the behaviours of its components, i.e. the buildings and occupants are considered in this study[151]. The urban energy demand is usually calculated as the sum of the energy demand (see **Table 3**) of each building or by using building archetypes that are representative of the building stock [152]. Three tools currently consider the geometry of every building and allow for yearly dynamic simulations with an hourly resolution of the urban energy demand: CitySim, UMI and CEA. To the best knowledge of the authors, there are no other tools able to simulate accurately and explicitly the power demand of urban buildings at the entire city scale with an hourly time step. This can be explained, at least partly, by the substantial computational cost required and also with regards to the data and information required to set up the tools.

CitySim supports the environmental design of urban master plans, by using the SUNtool [153] solver as a reduced dynamic thermal simulation platform. Compared to traditional building energy simulation tools, such as EnergyPlus [154], CitySim can quantify the energy demand at the urban scale, with a higher spatial resolution. It is, therefore, possible to explicitly simulate whole neighbourhoods or districts to predict the energy demand of buildings over a year for a considerable number of single buildings. It applies properties of shortwave and longwave which considers obstructions to sun and sky and reflections coming from the adjacent obstructions and uses them as input. The prediction of internal lightning rate and internal temperature are included and allow for accounting the occupants' behaviour. Citysim is highly compatible with Rhino tools, and Grasshopper interfaces are under development [155], in order to simplify the interconnection between the tool and architecture firms. The CIM was previously interconnected with the tool CitySim [130], in order to compute the microclimatic conditions within the urban setting, understanding the impact of the urban microclimate on the energy demand of buildings [81], [156] and the outdoor thermal comfort [128]. CitySim can also provide a computation of the natural ventilation potential and also accounts for the solar potential (solar heat gains, solar photovoltaic production,…).

The Sustainable Design Lab has developed an Urban modelling interface (UMI) at the Massachusetts Institute of Technology [157]. The goal of this Rhinoceros-based urban modelling design tool is to improve the efficiency of new and existing neighbourhoods regarding sustainability, in conjunction with operational energy use, daylighting, outdoor comfort and walkability. UMI is based on the Energy Plus engine and computes the energy demand of each building individually while considering the neighbouring building as shadowing obstacles.

The City Energy Analyst (CEA) model [158] was developed, with the scope of determining the spatiotemporal variability of the energy services in the future. This itself was a hybrid method, which contains the following four main phases; statistical model (1), analytical model (2), aggregation (3) as well as clustering and visualisation (4). The envelope model is based on electrical analogy obtained by discretisation of the wall in layers (usually one, two or three) characterised by specific thermal resistance and capacitance. The model was validated with a peer model and empirical data. This framework was created to analyse different urban scenarios by the energy, carbon emission, and financial point of view. Currently, a Grasshopper interface is under development.

Modelling urban morphology can be divided into three main approaches based on studying; possible configurations [159]; real-site configurations [160]; or both [152], [161]. Through applying any of these approaches, several studies have investigated the relationship between urban morphology and the energy demand and supply of buildings in urban areas, each considering some influencing parameters of urban morphology in terms of energy demand [162], [163] and wind assessment for ventilation [164]. Javanroodi et al. [33] provided a review of the available methods for simulating the thermal performance and ventilation potential of buildings in urban areas and introduce a novel approach to model and assess the impacts of urban morphology on the energy performance and ventilation potential of buildings in urban areas. Finally, several other tools and methodologies are under development, aiming to provide a comprehensive methodology to model urban buildings. Two examples are CESAR, focusing on bottom-up buildings stock modelling [165] or Urban Solve, to support the neighbourhood design at the masterplan stage [166]. Some recent studies have also used Energyplus as a tool to simulate the energy demand of buildings (at the individual scale [167] or at the urban scale [168]) to perform energy system optimisations.

*Table 3 : Selection of criterions used for energy demand/supply*

| Demand | Supply |
|---|---|
| • Heating load | • Natural Ventilation |
| • Cooling load | • Solar potential |

## Section 4.4: Energy systems

Assessment of energy infrastructure has been often focused on power generation. The primary motivation behind this has been the replacement of fossil fuel generation by using renewable energy technologies and thereby minimising the carbon impact [169]. Different computational tools have been developed to perform this specific task as reviewed by Lund et al. [170]. In addition to the computational tools available, some studies have proposed computational algorithms to conduct energy system optimisation [171]–[173]. The basis of the assessment has changed from simple economic analysis into eco-environmental assessment focusing on the carbon emissions during the recent past.

Furthermore, exergy efficiency or utilisation of renewable energy has also been considered recently. This makes it essential to develop design tools that can optimise more than one objective function during the optimisation process. A detailed review of different optimisation algorithms that can be used to optimise distributed energy systems has been presented in [174]. Both Pareto multi-objective optimisation and weighted-multi objective optimisation have been used to design urban energy systems considering both generations as well as distribution. A pool of criterions that can be considered in the process has been reviewed in [175]. Several limitations can be observed in these studies focusing on the optimisation:

- limitation to the boundaries of the energy system instead of considering the interactions int the infrastructure other than energy such as transportation, buildings, waste management, water supply etc:
- poor representation of uncertainties during the modelling, simulation, optimisation and assessment phases;
- limited to Pareto optimisation instead of extending it to the decision-making process;
- limited opportunities to bring experts having different backgrounds into the assessment process;
- poor justification for the specific criteria and preferences (such as a weight matrix during the decision-making process) selected for the assessment.

When considering these limitations, the first one relates to the energy nexus meaning that urban energy model should be further extended. However, decision making under uncertainty has yet to be included in the urban energy assessment process. The last three relates to the linking of energy infrastructure designing and decision-making process which requires major attention.

Extension of the urban energy model considering the nexus of water, food, transportation, agriculture has taken the attention recently focusing. More importantly, integrating building stock into urban energy structure has been widely discussed. In these instances, assessment of the energy efficiency of building stock and energy system has been performed independently [176]. However, their computational platforms have been developed in order to combine building performance simulation and energy system designing together [176], [177]. Nonetheless, buildings have been taken as standalone structures without considering the thermal interactions among them. Although such an approach can be used to improve the energy efficiency of buildings and energy systems, there are limitations in using such models for urban planning purposes. Shi et-al [178] highlighted the importance of considering energy interaction among buildings when optimising the energy systems. Schüler et-al [179] tried to optimise the energy system and urban form considering the thermal interactions among buildings and subsequently conduct a comprehensive assessment. However, the impact of urban climate is not considered in this study. As demonstrated in the previous section, the urban climate plays a vital role when it comes to heating and cooling demand primarily due to climate change [80]. A significant extension is required in the urban energy system model in order to consider the influence of urban climate. Perera et-al. [117] extended the boundaries of the urban energy system model to incorporate the influence of urban climate. They showed that the urban climate is having a considerable impact on both the design and operation of the energy system. Mauree et al., [80] looked at the impact of climate change on future energy demand. However, none of these studies has looked at both the influence of climate change on the building stock and subsequently on the energy system. Extreme climate conditions may take place frequently as a result of climate change which will notably influence the energy demand making it essential to have the energy system to become climate resilience. Therefore, the climate resilience of urban energy infrastructure is an essential aspect to be considered when designing resilient cities [180].

Extending the energy system optimisation process considering multi-criterion assessment and decision making is important to bring experts from different backgrounds into one table. Furthermore, this will enable to incorporate the inputs from different stakeholders of the city into the planning process. Different techniques such as Fuzzy-TOPSIS [171], [181], Analytical Hierarchical Process etc. have been used to consider multiple criteria during the decision-making process. As discussed previously, combining urban planning and energy system designing will lengthen the simulation and optimisation process. Incorporating decision-making into this will further extend the process [182]. Such wide-ranging processes will be difficult to implement and would be highly specific to the particular application. This makes it important to have a standard set of performance indicators that can be used to assess the energy sustainability of the urban planning process. Bringing climate resilience and adaptation needs to be major priorities when defining such a common set of performance indicators.

The criterions proposed can be classified into two blocks: (1) hard criterions based on the typical 3E (energy, economy and environmental) and soft criterions to present social aspects. Soft criterions are ambiguous in certain instances and highly case-specific and are hence not treated here. Major or hard criterions can be defined and evaluated using a mathematical model straight forward at the design phase. A list of major criterions that are already used to assess different cases which appeared in more than 100 recent publications can be found in Ref. [183]. We here provide a selection of frequently used criterions to assess distributed energy systems (see **Table 1**).

*Table 4 : Selection of criterions used for energy systems*

| Economy | Environmental impact | Energy |
|---|---|---|
| - Initial Capital Cost<br>- Net Present Value<br>- Cost of Energy<br>- Levelized Cost of energy | - Lifecycle CO2<br>- Renewable Energy Integration<br>- Normalized Capacity<br>- Annual percentage contribution in the generation | - Loss of Load Expectation<br>- Loss of Load Duration<br>- LOLF Loss of Load Frequency<br>- Loss of Load Probability<br>- Autonomy<br>- Utilisation / Waste of Renewable Energy<br>- Exergy / Energy efficiency |

## Section 5: Climate uncertainty

Over and above the points mentioned previously, major challenges exists in the climate adaptation of the built environment for the future climate due to the nature of climate and its stochastic behaviour, which induces large uncertainties in the assessment [34]. Future climate conditions are simulated by global climate models (GCMs) using different initial conditions, GHG emission scenarios or GHG concentration pathways, also known as Representative Concentration Pathways (RCPs)[184]. The spatial resolution of GCMs is quite coarse (around 100-300km$^2$) [1], and the direct use of their outputs is not recommended [185]. Therefore, the GCM data should be downscaled, using one of the two major approaches for downscaling: dynamical and statistical [36]. Many of the impact assessment studies (as the necessary action for climate adaptation), need hourly and even sub-hourly temporal resolution, especially those related to extreme conditions. The statistically downscaled data, such as morphed data [186], does not reflect future climate variations and extreme conditions and underestimates the impacts of climate change [187], [188]. Therefore, it is recommended to use dynamically downscaled weather data which are simulated by regional climate models (RCMs) and have suitable temporal (hourly to sub-hourly) and spatial resolutions (2.5km$^2$ and even less), respectively [189], [190]].

The generated weather data will be different depending on the selected GCM, RCM, emissions scenario, RCP and spatial resolution [191]. Therefore, it is not possible to plan climate change adaptation strategies based on a few numbers of climate scenarios [1]. It is also not possible to rely on short periods (days) and long periods (yearly or more) should be considered since the natural variability in the climate system makes the short term comparisons unreliable [192]. This means for having a proper impact assessment of climate change we need to deal with big datasets which makes the calculations expensive [191]. This brings up the importance of synthesising the right type of representative weather data sets that shorten the assessment while representing typical and extreme conditions and account for climate uncertainties.

One common approach for generating representative (past) weather data in energy and environmental studies is generating one typical year out of 30 years. Several techniques are available to create typical or reference weather files which have been reviewed in some works (e.g. [36], [193]). One well-known weather data type is the typical meteorological year (TMY) [184], which is based on selecting typical meteorological month (TMM) for each month using Finkelstein–Schafer (FS) statistics [184]. These types of average weather data set mostly represent only average/typical conditions and cannot take into account extreme conditions, resulting in significant underestimation or overestimation in calculating peaks and extremes [194]. Concerning future climate files, most of the available files are based on creating typical conditions using statistically downscaled GCM data (e.g. [187], [195]), neglecting future climate variations and anomalies and therefore extreme conditions. Some methods have been developed to consider extreme conditions, such as the proposed methods by Crawley et al. [196] and Nik [36]. Crawley et al. [196] created Extreme Meteorological Year (XMY) using four combinations of extremes : daily maximum, daily minimum, hourly maximum and hourly minimum for an initial set of variables of dry-bulb temperature, dew-point temperature, solar insolation, precipitation, relative humidity, and wind speed. Nik [36] proposed a method for generating representative future weather data sets out of RCMs, based on synthesising one typical and two extreme (cold and warm) data sets: Typical Downscaled Year (TDY), Extreme Cold Year (ECY) and Extreme Warm Year (EWY). The method has the advantage of simplifying the procedure for synthesising representative weather files while including extreme conditions and considering future climate uncertainties. The application of the method has been proven for several types of simulations and impact assessment studies [186], [197].

In addition to the uncertainties related to the climatic data and its implication on the energy demand and comfort, the reliability of the energy systems will also be significantly impacted. The potential of renewable energy generation coupled with the uncertainties related to climate change could have considerable consequences on the grid stability for example. Data-driven or machine learning approaches can be used to provide more reliable forecasting that could help the system to self-adapt [198], [199].

## Section 6: Integrated Workflows

In the previous sections, we have demonstrated that there was strong evidence on the interrelation between the urban fabric and texture, the urban energy demand, the outdoor thermal comfort and the energy system. Referring to the example of hot climates, it was shown that wind flow, induced by urban form design, plays an essential role in passive or active ventilation systems, helping to reduce the cooling load of buildings [200] and urban heat island [201] and to enhance the thermal comfort and thermal circulation around buildings. Furthermore, research works have pointed to an indirect relation between wind flow rate in urban canopies and average surface temperature in urban areas [31], [32] which can directly or indirectly affect the heat gain through external walls [53] and consequently the cooling load of building [80]. Thus, both thermal and wind flow characteristics of the urban fabric should be taken into account to design cooling load and ventilation strategies in an urban area that are prone to UHI phenomena. Additionally, buildings or energy systems that are designed today, will still be here in 30-100 years. This means that we also need to account for the climatic variability related to the urban space and also due to future climate change.

As underlined within the text, it is currently difficult to find a comprehensive methodology, or a tool, able to compute the outdoor environmental conditions, focusing on the energy demand of buildings, the energy systems and the pedestrian's thermal comfort. We have thus tried to regroup these processes in one framework to show the possibilities of building one ecosystem that could be used to design more sustainable urban areas. **Figure 4** represents a conceptual integrated workflow where the different elements are interrelated.

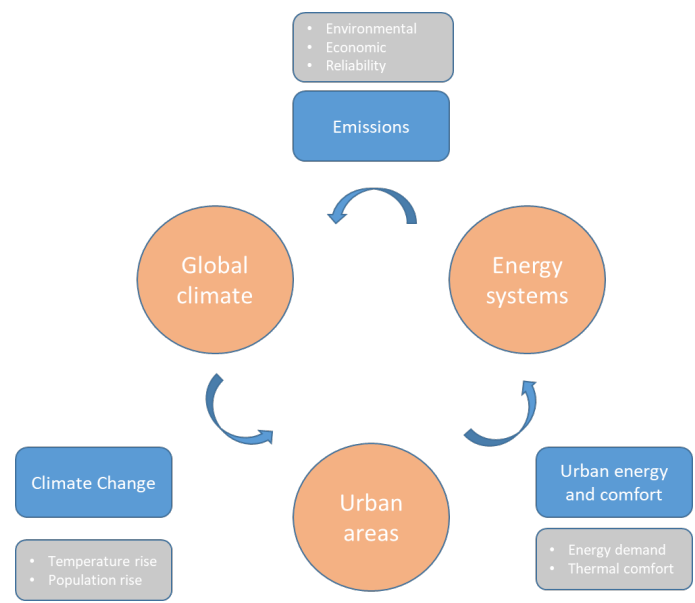

**Figure 4: Conceptual integrated workflow**

A few tools, such as CitySim or the City Energy Analyst, have been developed in recent years and provided an interesting systemic approach. As explained in Section 3, both models were developed as an urban modelling platform that includes integrated custom modules for modelling microclimatic effects, transient heat flow, plants and equipment as well as occupant presence and behaviour. CitySim has furthermore been extended with the coupling with CIM and the energy hub tool to design urban energy systems [117]. Table 5 underlines the tools that are already available and their capabilities. In particular, the possible connections and missing links in the various tools can be noted and calls for future addition of the missing features. Two parameters are nevertheless crucial in the evaluation of urban design: availability of yearly simulation and the spatial domain larger than a neighbourhood.

*Table 5 Capabilities of tools for evaluating the microclimate, the outdoor comfort and the energy demand and systems*

| Tool | Outdoor Thermal Comfort | Urban environmental conditions | Energy demand | Renewable productions | Energy systems | Yearly simulation | Neighbourhood / City scale simulation | References |
|---|---|---|---|---|---|---|---|---|
| **Microclimate** | | | | | | | | |
| ENVI-met | x | x | | | | | x | [140] |
| CIM | | x | | | | x | x | [120] |
| UWG | | x | | | | x | x | [122] |
| CAT | | x | | | | | x | [123] |
| SOLWEIG | x | x | | | | x | x | [141] |
| | | | | | | | | |
| **Comfort** | | | | | | | | |
| RayMan | x | x | | | | x | | [139] |
| OTC Model | x | | | | | x | | [135] |
| UTCI calculator | x | | | | | x | | [147] |
| | | | | | | | | |
| **Energy demand** | | | | | | | | |
| Ladybug and Honeybee | x | x | x | x | | x | x | [143] |
| CitySim | x | x | x | x | x | x | x | [137] |
| CESAR | | x | x | | | x | x | [165] |
| EnergyPlus | | | x | | | x | | [167] |
| UMI | | | x | | | x | x | [157] |
| CEA | | | x | x | x | x | x | [158] |
| Urban Solve | | | x | | | x | x | [166] |
| | | | | | | | | |
| **Energy System** | | | | | | | | |
| Homer | | | | x | x | x | x | [208] |
| Perera et al., | | | | x | x | x | x | [171] |
| Energy hub | | | | x | x | x | x | [173] |

It should be highlighted that geospatial information are nowadays readily available either from satellite or from municipalities. These significantly improves the way information about buildings and the urban areas are introduced in the models. Both CitySim and the CEA are connected to GIS-based software. GIS tools provide an opportunity to use already available 3D dataset (such as with CityGML) making it much easier to obtain rapidly usable inputs. Additionally, they can also be used to analyse and process outputs from simulations and provide powerful decision-making tools [202].

# Section 7: Conclusion
## Section 7.1 Major findings
Climate change and the urban microclimate, directly and indirectly, impact the outdoor thermal comfort, the energy demand in buildings and the energy systems. If the goal is to design more sustainable urban areas, based on the Sustainable Development Goals (SDG), in particular SDG #7 which relates to providing affordable and clean energy, #11 on sustainable cities and communities and #13 on climate action, it is critical to assess the interdependencies among these actions in the urban built environment and the energy systems.

As shown in **Figure 3** and from the review we have conducted, current studies have focused on the links between the individual elements mentioned. We have however shown that although there was an obvious strong link between each element, they also had to be considered together. We were able to reveal that an integrated framework was needed and that this should be addressed in the near future either from a research perspective or from a planning one.

As it has been demonstrated in the Sections 2 and 3, design, siting, orientation, layout, and outdoor spaces configurations, make use of solar gain and microclimatic conditions to minimise the need for buildings heating, cooling and lighting by conventional energy sources. Calibrating the access to sun, wind and light when possible and admitting or blocking resources is often performed at the scale of the building or its parts. What has received less attention, however, is the possibility of applying this approach to a system where buildings and outdoor spaces collaborate to define the microclimate.

To improve the urban microclimate and to achieve energy demand savings and temperated outdoor spaces, we clearly showed the need to analyse the urban system as a complete ecosystem with a complex metabolism. The following parameters and recommendations should, therefore, be considered in urban design:

(1) built form, density and type - to impact airflow, view of sun and sky, and exposed surface area;
(2) street canyon, width-to-height ratio and orientation - to control warming and cooling processes, thermal and visual comfort conditions, and pollution dispersal;
(3) building design - to influence building heat gains and losses, albedo and thermal capacity of external surfaces, and use of transitional spaces;
(4) urban materials and surfaces finish - to influence absorption, heat storage, and emissivity;
(5) green and blue infrastructures - to facilitate evaporative cooling processes on building surfaces and/or in open spaces;
(6) traffic reduction, diversion, and rerouting - to reduce air and noise pollution and heat discharge;
(7) energy systems – to integrate more renewable energies and to decrease the carbon footprint of urban areas;
(8) climate change – to take into account the future challenges related to climatological extremes in the urban environment.

This is very much in line with what was referred to as the "Nexus approaches" by Bai et al.,[203]. Darchen and Searle [204] have also previously demonstrated that urban sustainability should be addressed in all the "three arenas : equity and justice sustainability, environmental sustainability and economic sustainability". In our study, we focused mostly on the environmental sustainability but showed how it was strictly linked to the well-being of the inhabitants and how this can also have an economical impact. It is clear that a holistic and comprehensive framework that links buildings operations, outdoor spaces and outdoor climate is necessary, to create a positive loop in the urban design. Only stronger national and international energy regulations for urban design to improve the energy sustainability at the city and national level will allow cities and urban planners to move in this direction. The urban climatic map should be integrated in the city design, as a new and smart instrument that can be used for developing adaptation strategies to face future climate change within the urban environment. Additionally, international energy policy should be developed, regulating the energy infrastructures, as an example an improved European energy hub, able to store, share and redistribute the renewable energy with the use of big data and through machine learning techniques.

## Section 7.2 Future perspectives

One of the challenges that remain to be addressed is the fault detection in the system, in particular with possibilities of variations with high amplitude for example during heat waves or climatological events. With the increased use of smart devices [205] and the collection of massive data, it can, however, be expected that techniques using data-driven or machine learning approaches will be used to improve the forecasting capabilities and/or to locate faults faster [206]. It can also be noted that with the development of smart objects and devices, data collection and availability will provide more insights in user behaviour and could be used to derive data-driven approaches in order to design and built more sustainable and resilient cities. Understanding and materialising the dynamics of cities with Internet of Things and creating smart interconnected cities will create more adaptable and reactive environment that can act and react as function of the citizens and empower these latter more. This could bridge the existing gaps that were mentionned by Kavgic et al., [207].

Over and above the points mentioned before, the uncertainty related to climatic projections for the future as well as the methodology to downscale the data, is one of the major challenges that need to be addressed to support urban planning processes related to climate change adaptation. Furthermore, the bottom-up approach from the building to the city scale, can be integrated into the assessment at the national scale to derive trans-national policies regarding climate change mitigation. As pointed out by Moriaty [208], the complete transition of energy systems to renewable energy will not be possible unless the energy efficiency of urban areas is addressed. This applies in particular to urban areas, where building envelopes, walls and roofs, can be used to decrease the energy footprint of the buildings and to transform the solar irradiation into useful solar heat gains.

Few limitations to this study can also be raised. We did not look at the life cycle of the buildings which should be taken into account especially at the city scale [209]. Additional, we have not addresssed the social aspects of urban design in particular on the acceptability of the energy transition or the implementation of energy efficiency measures at the urban scale. Future studies should also address these aspects along with health-related issues including air pollutants emissions and transportation in urban areas.

**Acknowledgements:** This research project has been financially supported by the Swiss Innovation Agency Innosuisse and is part of the Swiss Competence Center for Energy Research SCCER FEEB&D. The article was developed with the support of the COST Action CA16114 'RESTORE: Rethinking Sustainability towards a Regenerative Economy'.